\documentclass{emulateapj}
\usepackage{apjfonts}
\usepackage{epsfig,color}
\usepackage{mathrsfs}
\usepackage{ulem}

\def\bhm{M_{\bullet}}

\def\dotm{\dot{m}}
\def\ergs{\rm erg~s^{-1}}
\def\fGR{f_{_{\rm GR}}(r,a)}

\def\rblr{R_{\rm BLR}}
\def\rblrl{\rblr-L_{5100}}
\def\rsqrt{\sqrt{r}}
\def\rms{r_{\rm ms}}
\def\sigmaT{\sigma_{\rm T}}
\def\sunm{M_{\odot}}

\journalinfo{The Astrophysics Journal Letters 2014 (in press)}
\slugcomment{Received 2014 July 8;  accepted 2014 August 6}

\begin{document}

\title{A New Approach to Constrain Black Hole Spins in Active Galaxies 
Using Optical Reverberation Mapping}

\author
{Jian-Min Wang\altaffilmark{1,2},
Pu Du\altaffilmark{1},
Yan-Rong Li\altaffilmark{1},
Luis C. Ho\altaffilmark{3,4},
Chen Hu\altaffilmark{1}, and
Jin-Ming Bai\altaffilmark{5}
}

\altaffiltext{1}
{Key Laboratory for Particle Astrophysics, Institute of High Energy Physics,
Chinese Academy of Sciences, 19B Yuquan Road, Beijing 100049, China.}

\altaffiltext{2}
{National Astronomical Observatories of China, Chinese Academy of Sciences,
 20A Datun Road, Beijing 100020, China}

\altaffiltext{3}
{Kavli Institute for Astronomy and Astrophysics, Peking University, Beijing 100875, China} 

\altaffiltext{4}
{Department of Astronomy, Peking University, Beijing 100875, China} 

\altaffiltext{5}{Yunnan Observatory, Chinese Academy of Sciences, Kunming 650011, China}

\begin{abstract}
A tight relation between the size of the broad-line region (BLR) and optical 
luminosity has been established in about 50 active galactic nuclei studied 
through reverberation mapping of the broad H$\beta$ emission line. The 
$\rblr-L$ relation 
arises from simple photoionization considerations.  Using a general 
relativistic model of an optically thick, geometrically thin accretion disk, 
we show that the ionizing luminosity jointly depends on black hole mass, 
accretion rate, and spin.  The non-monotonic relation between the ionizing 
and optical luminosity gives rise to a complicated relation between the BLR 
size and the optical luminosity. We show that the reverberation lag of 
H$\beta$ to the varying continuum depends very sensitively to black hole 
spin.  For retrograde spins, the disk is so cold that there is a deficit of 
ionizing photons in the BLR, resulting in shrinkage of the hydrogen ionization 
front with increasing optical luminosity, and hence shortened H$\beta$ lags.  
This effect is specially striking for luminous quasars undergoing retrograde 
accretion, manifesting in strong deviations from the canonical $\rblr-L$ 
relation.  This could lead to a method to estimate black hole spins of 
quasars and to study their cosmic evolution.  At the same time, the small 
scatter of the observed $\rblr-L$ relation for the current sample of 
reverberation-mapped active galaxies implies that the majority of these 
sources have rapidly spinning black holes.

\end{abstract}


\section{Introduction}
The broad emission lines from quasars and active galactic nuclei (AGNs) 
largely reflect virial motions of broad-line region (BLR) clouds governed 
by the gravitational potential of the central supermassive black hole (BH). 
The BLR clouds are mainly photoionized by the continuum emission from the 
accretion disk (e.g., Osterbrock \& Mathews 1986; Ho 2008), and the detailed 
emission-line spectra depend, at least in part, on the spectral energy 
distribution (SED) of the disk.  The SED, in turn, varies as a function of 
BH mass, spin, and accretion rate.

Reverberation mapping (RM; Blandford \& McKee 1982) of $\sim 50$ AGNs has 
discovered a very tight correlation between BLR size, as measured from the 
H$\beta$ lag in response to the varying continuum, and the optical luminosity 
(popularly measured at 5100 \AA; Kaspi et al. 2000; Bentz et al. 2013). This 
well-known relation can be described as
$\rblr\approx 36.3~ L_{44}^{\gamma}~{\rm ld}$, 
where $\gamma = 0.546^{+0.027}_{-0.027}$ and $L_{44}$ is the 5100 \AA\ 
luminosity $L_{5100}$ in units of $10^{44}\, \ergs$.  This is a natural 
consequence of photoionization of BLR clouds.
Defining the ionisation parameter as
$U=Q_{\rm ion}/4\pi R_{\rm H\beta}^2c n_e$, with $n_e$ the electron density,
$Q_{\rm ion}$ the rate of ionizing photons, and $c$ the speed of light, this 
empirical $\rblr-L$ relation can be explained if the photoionized clouds obey 
the condition $n_eU\sim 10^{10}$ (Negrete et al. 2013). Alternatively, 
sublimation of dust particles and energy balance between heating and radiation 
loss in the BLR produces a similar scaling relation between $\rblr$ and $L$ 
(Netzer \& Laor 1993).  In either case, this observational relation sets 
strong constraints on the physical properties of the BLR.  It is of interest 
to explore how the $\rblr-L$ is affected by the ionizing SED.

In this Letter, we show that ionizing luminosity $L_{\rm ion}$ is a 
complicated function of BH mass, spin, and accretion rate. It can be much 
fainter than $L_{5100}$, and thus the H$\beta$-emitting region can be much 
smaller than that predicted by the $\rblrl$ relation, which is based on 
$L_{5100}$. We emphasize the influence of BH mass and spin on the
SED, and thus on H$\beta$ reverberation. 
The sensitivity of H$\beta$ lags to BH spin provides a promising method to 
estimate spins from RM observations.

\section{Ionizing Luminosity from Accretion Disks}
\subsection{Emissions from Accretion Disks}
We employ the general relativistic version of the Shakura-Sunyaev (1973) disk 
(Page \& Thorne 1974). General relativistic effects are important because the
ionizing photons are emitted from regions very close to the innermost last 
stable orbit ($\rms$), especially for extremely massive BHs 
($\sim10^9\,\sunm$).  We define the dimensionless accretion rate as
$\dotm=0.1\dot{M}_{\bullet}c^2/L_{\rm Edd}$, 
where $\dot{M}_{\bullet}$ is the accretion rate, $L_{\rm Edd}=4\pi G\bhm m_pc/\sigmaT$ 
is the Eddington luminosity, $\bhm$ is the BH mass, $\sigmaT$ is the 
Thompson cross section, and $G$ is the gravitation constant.  Disks with
accretion rates $\dotm \approx 0.01-0.3$ are in the Shakura-Sunyaev regime, 
characterized by being geometrically thin and optically thick. Above this 
regime in $\dotm$, accretion onto BHs is described by slim disks (Abramowicz 
et al. 1988). Wang et al. (2014) recently discuss the influence of slim disks 
on the BLR.   This paper is devoted to thin disks. The radiation flux from the 
disks is given by
\begin{equation}\begin{array}{lll}
F¨&=&\displaystyle\frac{3m_pc^5}{2\sigmaT G\bhm}\frac{\dotm}{r^3}\fGR\\ \\
    &=&6.86\times 10^{17}~\dotm_{0.1} M_9^{-1}r^{-3}\fGR
                ~{\rm erg~s^{-1}~cm^{-2}},\end{array}
\end{equation}
where $\dotm_{0.1}=\dotm/0.1$,
$a$ is the specific angular momentum of the BH, $M_9=\bhm/10^9\,\sunm$, $r=R/R_{\rm g}$ 
is the dimensionless radius, and $R_{\rm g}=G\bhm/c^2=1.48\times 10^{14}~M_9$ cm. Here,
$\fGR$ is a factor that includes general relativistic effects (Page \& Thorne 1974):
\begin{eqnarray}
\fGR   &=&\displaystyle{\frac{1}{{\cal C}\rsqrt} \left\{\rsqrt-\sqrt{\rms}-
          \frac{3a}{4}\ln\frac{r}{r_{\rm ms}}
          -3\left[\frac{(x_1-a)^2}{x_1(x_1-x_2)(x_1-x_3)}\right.\right. \nonumber}\\
       & &  \displaystyle{\left.\left.\times \ln\frac{\rsqrt-x_1}{\sqrt{\rms}-x_1}
         +\frac{(x_2-a)^2}{x_2(x_2-x_1)(x_2-x_3)}\ln\frac{\rsqrt-x_2}{\sqrt{\rms}-x_2}\right.\right.\nonumber}\\
       & &  \displaystyle{\left.\left.         
         +\frac{(x_3-a)^2}{x_3(x_3-x_1)(x_3-x_2)}\ln\frac{\rsqrt-x_3}{\sqrt{\rms}-x_3}\right]\right\}},
\end{eqnarray}
where ${\cal C}=1-3/r+2a/r^{3/2}$, $x_1=2\cos[(\Theta_a-\pi)/3]$,
$x_2=2\cos[(\Theta_a+\pi)/3]$, $x_3=-2\cos(\Theta_a/3)$, and $\Theta_a=\arccos a$. 
Here $\rms$ is the last stable orbit, and can be expressed by
$\rms=[3+Z_2\mp \sqrt{(3-Z_1)(3+Z_1+2Z_2)}]$,
$Z_1=1+(1-a^2)^{1/3}\left[(1+a)^{1/3}+(1-a)^{1/3}\right]$ and $Z_2=\sqrt{3a^2+Z_1^2}$.
Figure 1 shows the function $r^{-3}\fGR$, which is related to the effective temperature 
of the disk, for different spins. We find that the peak of the dissipation rate is sensitive 
to $a$, lending us an opportunity to constrain the spin.

\begin{figure}[t!]
\begin{center}
\includegraphics[angle=-90,width=0.45\textwidth]{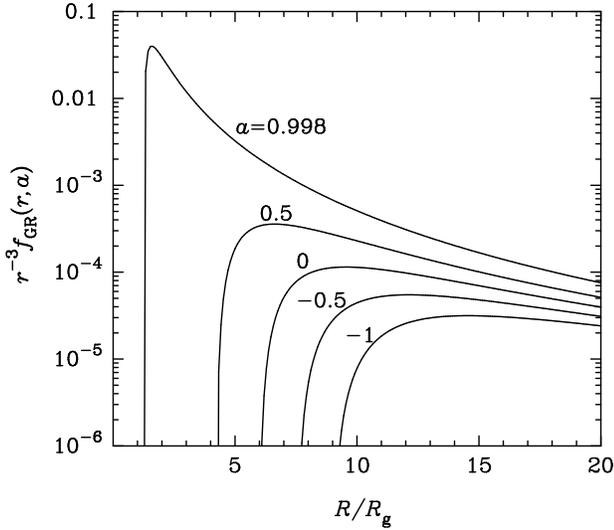}
\end{center}
\caption{\footnotesize  The function $r^{-3}\fGR$, for $a=0.998$ to $a=-1$. Two characteristics 
can be seen: (1) the last stable inner radii are different for $a=0.998$ and $a=-1$, and hence 
the total dissipated energy and SED are very different; (2) the function tends to $r^{-3}$ when 
$r\rightarrow \infty$, which leads to similar optical radiation (e.g., at 5100 \AA) for $a=\pm 1$.
}
\label{fig1}
\end{figure}

\begin{figure}[t!]
\begin{center}
\includegraphics[angle=-90,width=0.4\textwidth]{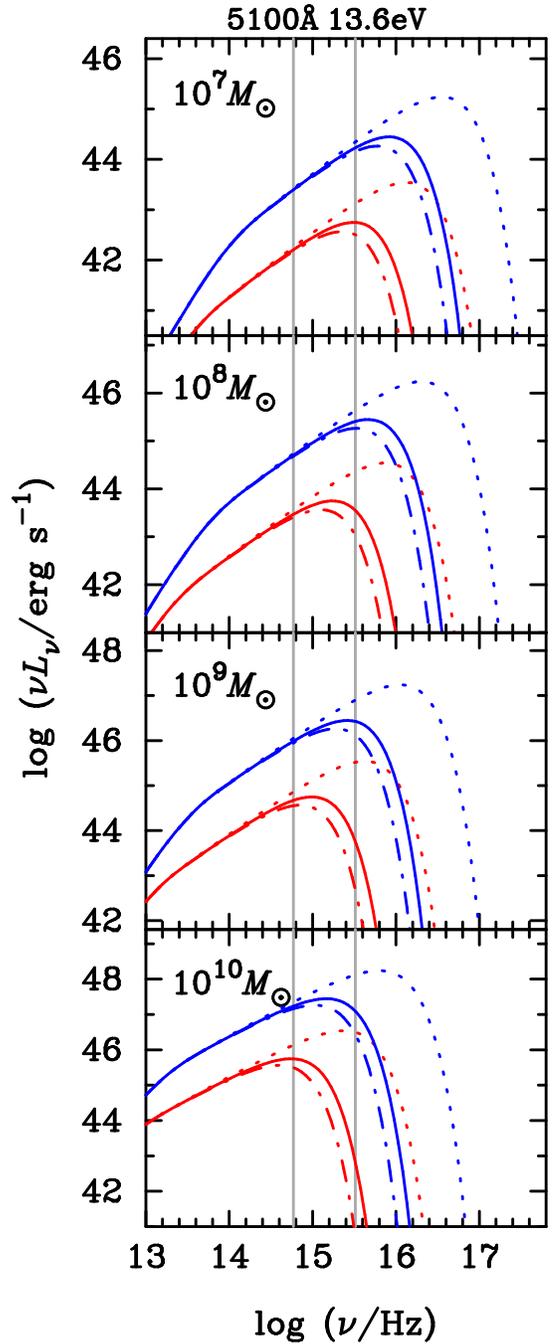}
\end{center}
\caption{\footnotesize The SEDs of accretion disks and their dependence on BH mass, 
accretion rate, and BH spin. The blue and red curves are for $\dotm=0.5$ and 0.01, 
and the dashed-dotted, solid, and dotted lines are for spin $a=-1,0,$ and 0.998, 
respectively. The two grey vertical lines mark 5100\AA\, and 13.6 eV.
}
\label{fig1}
\end{figure}

The SED is governed by the effective temperature distribution 
of the disk, which follows from  $T_{\rm eff}=\left[F(r)/\sigma_{\rm SB}\right]^{1/4}$, where 
$\sigma_{\rm SB}$ is the Stefan-Boltzmann constant,
\begin{equation}
T_{\rm eff}=3.32\times 10^5~\dotm_{0.1}^{1/4}M_9^{-1/4}r^{-3/4}f_{\rm R}^{1/4}~{\rm K},
\end{equation}
and can be obtained by integrating the entire disk from $\rms$ to infinity,
\begin{equation}
L_{\nu}=32\pi^2\cos i~ R_{\rm g}^2\int_{\rms}^{\infty}\frac{2h}{c^2}
        \frac{\nu^3rdr}{e^{h\nu/kT_{\rm eff}¨}-1},ÊÊ
\end{equation}
where $h$ and $k$ are the Planck and Boltzmann constants, respectively, and $i$ is the
inclination of the disk with respect to the line of sight. For a given wavelength 
$\lambda=c/\nu$, we have the luminosity $L_{\lambda}=\nu L_{\nu}$, which is used 
for the empirical correlation with the BLR size. Integration of Equation (4) 
yields the approximate functional form $L_{\nu} \propto \nu^{1/3}$.
For convenience, we set $\cos i=1$.

\begin{figure*}[t!]
\begin{center}
\includegraphics[angle=-90,width=0.95\textwidth]{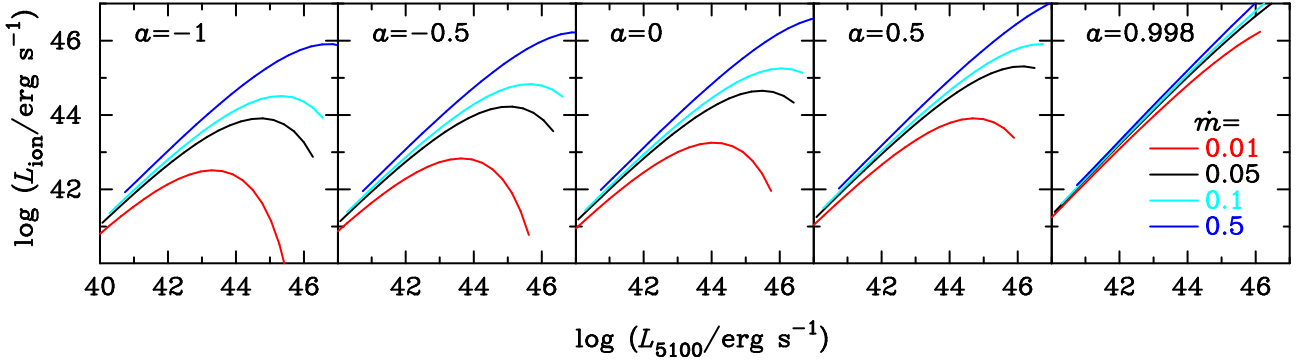}
\end{center}
\caption{\footnotesize Relation between the hydrogen ionizing luminosity and the 
luminosity at 5100 \AA.  Panels from left to right are for different values of BH 
spin, from $a=-1.0$ to $a=0.998$. Each panel plots four colour lines for different
accretion rates. Given $\dot{m}$ and $a$, $L_{5100}$ is a function of the 
BH mass.}
\label{fig2}
\end{figure*}

Figure 2 shows spectra from accretion disks with different parameters. We keep the 
accretion rates in the Shakura-Sunyaev regime in order to avoid the self-shadowing 
effects of the disk itself (Wang et al. 2014). 
For the same $\dotm$, retrograde spins produce much softer SEDs than prograde 
spins.  For each disk with ($\bhm,\dotm$), we plot spins of $a=-1$, 0, and 
0.998.  We find the following. (1) The luminosity at 5100 \AA, 
$L_{5100}$, is always insensitive to spin for $\dot{m}=0.5$ to 0.01,
except for extremely large BHs with $\bhm\gtrsim 10^{10}\sunm$.
(2) For a given $\dot{m}$, the SED is very sensitive to $a$, except for values 
of $a=0$ or $a=-1$. (3) For $\bhm\lesssim 10^8\sunm$ and higher accretion rate 
($\dot{m}\gtrsim 0.5$), the hydrogen ionizing luminosity depends very little on 
spin ($\Delta \log L_{\rm 13.6eV}\lesssim 0.5$ dex for $a=0.998$ and 
$a=-1$); thus, the present approach of constraining spins is not effective for 
less massive BHs (see discussion in \S2.2).  

\subsection{Photoionizing Luminosity}
The ionizing portion of the SED is sensitive to ($\bhm,\dotm,a$).  Integration 
of the SED above $\epsilon_1=13.6$ eV yields the ionizing luminosity.  To 
estimate the number of ionizing photons, we need to know the photoionization 
cross section of hydrogen (e.g., Osterbrock 1989).  For $h\nu\ge \epsilon_1$,
\begin{equation}
\sigma_{\nu}=6.3\times 10^{-18}\left(\frac{\epsilon_1}{h\nu}\right)^4
           \frac{e^{4-[(4\tan^{-1}\epsilon)/\epsilon]}}{1-e^{-2\pi/\epsilon}}~{\rm cm^{-2}},
\end{equation}
where $\epsilon=(h\nu/\epsilon_1-1)^{1/2}$, $h$ is the Planck constant; otherwise, 
$\sigma_{\nu}=0$ when $h\nu<\epsilon_1$.  The total luminosity to ionise hydrogen is 
\begin{equation}
L_{\rm ion}=\frac{\int_{\epsilon_1}^{\infty} \nu L_{\nu}\sigma_{\nu}d\nu}
            {\int_{\epsilon_1}^{\infty} \sigma_{\nu} d\nu}.
\end{equation}
The observed H$\beta$ luminosity is given by
$L_{\rm H\beta}=(\Delta \Omega/4\pi)(\epsilon_{_{\rm H\beta}}/\alpha_{\rm B})Q_{\rm ion}({\rm H^0})$,
where $\Delta \Omega$ is the solid angle subtended by the BLR, $\epsilon_{_{\rm H\beta}}$ is the 
recombination-line emission coefficient of H$\beta$, $\alpha_{_{\rm B}}$ is the radiative recombination 
coefficient for hydrogen quantum levels $n=2$ and higher, $Q_{\rm ion}({\rm H^0})$ is the ionizing
photon rate. 
In principle, the H${\beta}$ luminosity ($L_{\rm H\beta}$) emitted from the 
photoionized clouds in the BLR is proportional to $L_{\rm ion}$, but it also 
depends on some details of the clouds and the BLR. The $L_{\rm H\beta}-L_{5100}$ 
relation can be used to observationally test photoionization processes.

\begin{figure*}[t!]
\begin{center}
\includegraphics[angle=-90,width=0.95\textwidth]{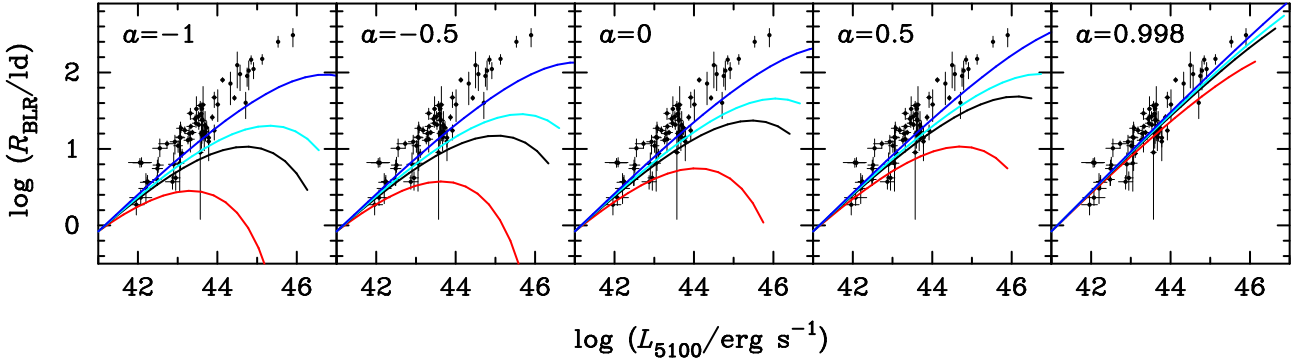}
\end{center}
\caption{\footnotesize  Relation between the BLR lag and the luminosity at 5100 \AA.  
Panels from left to right are for different values of BH spin, from $a=-1.0$ to $a=0.998$. 
Each panel plots four colour lines for different accretion rates. The normalization constants, 
for $\dotm=(0.01,0.05,0.1,0.5)$, are as follows:
$a=-1$: $R_0=(18.4,12.0,10.6,8.4)$ ld; 
$a=-0.5$: $R_0=(16.3,11.2,10.0,8.2)$ ld;
$a=0$: $R_0=(14.2,10.3,9.4,7.9)$ ld;
$a=0.5$: $R_0=(12.0,9.3,8.6,7.5)$ ld;
and $a=0.998$: $R_0=(8.3,7.4,7.2,6.6)$ ld.}
\label{fig4}
\end{figure*}

Figure 3 shows predicted $L_{\rm ion}-L_{5100}$ relations for disks with
$a=(-1,-0.5,0,0.5,0.998)$ and $\dotm=(0.01,0.1,0.5$). The ratio 
$\ell=L_{\rm ion}/L_{5100}$ is a complicated function of ($\bhm,\dotm, a$).  
For a given $a$ and $\dotm$, $L_{5100}$ increases with BH mass, approximately 
as $L_{5100}\propto \bhm^{1/3}$ for a Newtonian potential. For the case of 
$a=-1$ (left panel of Fig. 3), we find that $L_{\rm ion}$ linearly increases  
with $L_{5100}$ in logarithmic space until a critical luminosity $L_{5100}^c$. 
Above $L_{5100}^c$, $L_{\rm ion}$ decreases exponentially with $L_{5100}$. 
This means that $L_{\rm ion}$ does not follow a linear relation with the 
observed optical luminosity. $L_{5100}^c$ is jointly determined by $\dotm$ and 
$a$, increasing with increasing $a$.  For extreme prograde spins, the 
$L_{\rm ion}-L_{5100}$ relation remains linear for $\dotm\gtrsim 0.01$ up to 
$L_{5100}\approx 10^{47}\,\ergs$, which corresponds to the most massive 
BHs in quasars ($\sim 10^{10}\,\sunm$). The non-monotonic connection between 
$L_{\rm ion}$ and $L_{5100}$ has strong observational implications for the 
BLR as well as BH spins.

We note that the $L_{\rm ion}-L_{5100}$ relation discussed here is closely 
related to the well-known correlation between Balmer line luminosity and 
optical continuum, which is usually taken as evidence for photoionization 
(e.g., Shuder 1981; Greene \& Ho 2005). However, for the comparison between
H$\beta$ and 5100 \AA\ continuum, the rms scatter of either the 
luminosity-luminosity or flux-flux correlations is still $\sim 0.2-0.3$ dex, 
depending on the sample and method of spectral analysis (Greene \& Ho 2005; 
Hu et al. 2008; Shen et al. 2011).  The scatter might reflect
the range in BLR covering factors of individual objects, or, alternatively, 
variations in $\bhm$, $\dotm$, and $a$.  

The AGN continuum is composed of optical-UV blackbody radiation 
from the cold disk (the ``big blue bump'') and Comptonization of the disk 
photons into X-rays by a hot corona above the disk (e.g., Haardt \& Maraschi 
1991). The fraction ($f_c$) of the gravitational energy dissipated in the hot 
corona is anti-correlated with $\dot{m}$ (see Fig. 1b in  Wang et al. 2004), 
but it remains the largest source of uncertainty in the disk-corona system 
(e.g., Svensson \& Zdziarski 1994).  In principle, $f_c$ should be determined 
by the details of the accretion physics, such as the radial motion of the 
accretion flow, the magnetic field, and its reconnection (e.g., Merloni \& 
Fabian 2002; Wang et al. 2004; Cao 2009; Done et al. 2013; Uzdensky 2013).  
Observations show that the maximum value of $f_c$ is $\lesssim 0.5$ (between 
2 and 150 keV) for a Shakura-Sunyaev disk (e.g., Fig. 1 in Yang et al. 2007; 
Cao 2009), and variations in $f_c$ can change the ionizing luminosity by 
$\sim$0.3 dex. By contrast, Figure 2 shows 
that variations in $a$ (from $a=0.998$ to $-1$) causes $L_{\rm ion}$ to vary 
by $\gtrsim 0.5-2$ dex for $\dot{m}\lesssim 0.5$ and 
$\bhm=10^8-10^{10}\,\sunm$ (except for $\bhm\lesssim 10^8\,\sunm$; see Fig. 2), 
respectively.  The spin strongly influences the ionizing luminosity,
especially when $\dot{m}$ is decreased.
Therefore, the current approach does not apply to less massive 
BHs with high accretion rates, such as narrow-line Seyfert 1 galaxies.
Hot electron scattering of the disk surface may additionally modify the SED, 
but this influence does not exceed a factor of 2 (e.g., Czerny \& Elvis 1987) 
and is neglected in the present paper. 
Here we focus on first-order constraints on BH spin that can be derived from 
H$\beta$ reverberation. More detailed measurements of the BH spin will require 
a need more careful treatment of the disk-corona radiation and determination 
of the parameter $n_eU$ from photoinization calculations.

\begin{figure*}[t!]
\begin{center}
\includegraphics[angle=-90,width=0.95\textwidth]{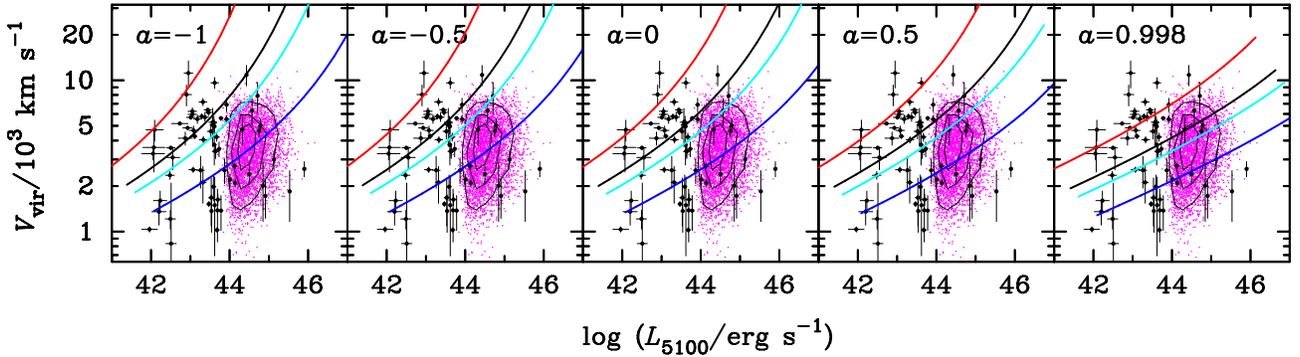}
\end{center}
\caption{\footnotesize The distribution of virial velocities for the BLR clouds 
as a function of optical continuum luminosity.  Panels from left to right are for 
different values of BH spin, from $a=-1.0$ to $a=0.998$. Each panel plots four 
curves for different $\dot{m}$ (corresponding to the same color).  The black 
points are the RM AGNs, the magenta points SDSS quasars from Hu et al. (2008).
The contours mark 20\%, 50\%, and 80\% of the sample. The regions of the diagram 
not occupied by the quasar sample may represent massive BHs and accretion rates 
$\dot{m}\gtrsim 0.5$, where the present approach does not apply.}
\label{fig5}
\end{figure*}

\subsection{Size-luminosity Relation}
The strong empirical correlation $\rblr\approx 33~L_{44}^{\gamma}$ ld (Bentz 
et al. 2013) is in good agreement with the constant-$n_eU$ model. $L_{5100}$ can 
be converted into $L_{\rm ion}$ for a given accretion disk model. We calculate 
the $\rblr-L_{5100}$ relation by inserting $L_{\rm ion}=\ell L_{5100}$ into 
the constant-$U$ model. Supposing that 
$\rblr\propto L_{\rm ion}^{\gamma}=\ell^{\gamma}L_{5100}^{\gamma}$, we have 
$\rblr=R_0\ell^{\gamma}L_{44}^{\gamma}$, where $R_0=36.3/\ell_0^{\gamma}$ 
is the normalization, and $\ell_0$ is $\ell$ obtained by setting $L_{5100}=10^{41}\,\ergs$ 
for given $a$ and $\dot{m}$.

Figure 4 shows the photoionzation-based model of the variation of BLR size 
with $L_{5100}$. The normalization is determined by the observed relation. The 
data of points in the plot are taken from Bentz et al. (2013). For retrograde 
spins ($a=-1$), the predicted time lags are generally shorter than the 
observed values unless the accretion rates are very high (up to the 
super-Eddington regime of slim disks with $\dotm\gtrsim 1.0$). Increasing the 
BH spin increases the lags and brings them into better agreement with the 
observed values. This is a consequence of the increased production of ionizing 
photons for larger values of $\dotm$ and $a$, which leads to expansion of the 
ionizing front of BLR clouds and thus longer lags.

BHs with spins $a\le 0$ lead to BLR sizes much smaller than those observed in
the $\rblrl$ relation. The sizes can be increased by elevating the accretion 
rates in the Shakura-Sunyaev regime, but they are still systematically too 
small. The only way to reconcile this discrepancy with the observed $\rblrl$ 
relation is to increase the BH spin. As described in Section 3.1, there are 
five RM AGNs with BH spins determined through {\it Suzaku}\ observations.  
They are all fast-rotating BHs, and they are all consistent with the last 
panel ($a = 0.998$) in Figure 4.  We deduce that most of the $\sim 50$ RM AGNs 
are rotating maximally.

This work shows that BHs undergoing retrograde accretion can produce H$\beta$
lags much shorter than those observed on the $\rblrl$ relation. Self-shadowing 
effects from a slim disk (Wang et al. 2014), if present, further shorten the 
lags.  Since slim disks emit a distinctive SED (Wang et al. 2014), the two 
effects can, in principle, be disentangled.  The  $\rblrl$ relation of the 
current sample of RM AGNs already has a remarkably tight of 0.13 dex (Bentz et 
al. 2013). Some of the scatter surely must arise from variations in accretion 
rate and physical conditions of the BLR.  Thus, barring strong selection 
effects of some kind, the tight observed $\rblrl$ relation strongly suggests 
that the current sample of RM AGNs statistically have high BH spins.  By the 
same token, future, larger RM samples provide an opportunity to discover BHs 
with low or even retrograde spins by identifying outliers in the 
$\rblrl$ relation.

\subsection{FWHM and Luminosity Plane}
The virial velocity of the clouds in the BLR is given by 
\begin{equation}
V_{\rm vir}=c/\sqrt{r_{_{\rm BLR}}}=9.9\times 10^3~\ell_{10}^{-0.27}M_8^{1/2}L_{44}^{-0.27}~{\rm km~s^{-1}},
\end{equation}
where $r_{_{\rm BLR}}=\rblr/R_{\rm g}$ is the BLR size in units of $R_{\rm g}$, 
$\ell_{10}=\ell/10$, and $M_8=\bhm/10^8\,\sunm$. Since
$\ell$ is a function of ($\bhm,\dotm,a$), the plane formed by the two observables ${\rm FWHM}$ 
and $L_{5100}$ can deliver information on the spin. 

We plot the RM AGNs as red points in Figure 5; they are quite homogeneously 
distributed in the ${\rm FWHM}-L_{5100}$ plane. As shown by the panels, the 
distribution of RM AGNs is mostly covered by model lines that cover 
$\dotm\in(0.1,0.5)$ and $a=0.998$. We also show the quasar sample from Hu 
et al. (2008).
Interestingly most 
quasars occupy the region of parameter space $\dotm\in(0.1,0.5)$, in general
agreement with the observed Eddington ratio distribution of Sloan Digital 
Sky Survey quasars (e.g., Shen et al. 2011), and $a=0.5$.

\section{Discussion and Summary}

Measurements of BH spins are elusive in AGNs. 
The iron K$\alpha$ line profile, broadened and skewed by the strong gravity
around the BH, is usually employed to estimate the spin of the hole (e.g.,
Fabian et al. 2000; Risaliti et al. 2013; see Reynolds 2013 for review).  A
significant fraction of the AGNs with robust constraints on their spin from
X-ray observations with {\it XMM-Newton}, {\it Suzaku}, and {\it NuSTAR}\
have spin parameters $a>0.5$ (Reynolds 2013).  Among these, five have been
monitored by RM. {\it Suzaku} observations (Walton et al. 2013) report high
spins: $a = 0.86$ (Mrk 509), 0.83 (Mrk 335), $> 0.64$ (Fairall 9), $> 0.81$
(Ark 120), and $> 0.99$ (Ark 110).  In principle, BH spins for AGNs can also
be obtained from fitting the continuum with disk models (Czerny et al. 2011;
Done et al. 2013), but it is often challenging to measure the
full SED simultaneously.  More statistical techniques rely on estimating the 
average radiative efficiency of AGN populations, and using this information 
to infer the implied BH spin and its evolution with redshift (e.g., Wang et 
al. 2009; Li et al. 2012; Volonteri et al. 2013).

This paper offers a new approach to measure BH spins in AGNs using RM of the 
H$\beta$ line.  Studying the BLR size-luminosity relation based on a standard 
accretion disk model, we find that the BLR size is very sensitive to the spin 
of the BH.  BHs with low spins or retrograde accretion exhibit significantly 
shorter H$\beta$ lags in response to the varying optical continuum. In 
practice, our technique can be applied as follows.  Using RM observations, we
can derive the BH mass $\bhm=f_{\rm BLR}\rblr V_{\rm H\beta}^2/G$, where the 
virial factor $f_{\rm BLR}$ can be obtained from calibrations against the 
local scaling relations between BH mass and bulge properties (e.g., Onken et 
al.  2004; Ho \& Kim 2014) or from more sophisticated dynamical models of the 
BLR (Pancoast et al. 2011; Li et al. 2013).  As shown in Figure 2, the optical 
part of the disk continuum is insensitive to the spin, allowing us to 
estimate the accretion rates $\dot{M}_{\bullet}$. Then, invoking Figure 4, we 
can estimate the spin from the $\rblrl$ relation.  Applying this technique to 
the currently available sample of $\sim 50$ RM AGNs, we find that the majority
of them are consistent with having fast-rotating BHs.

Apart from the $\rblrl$ relation, we also discuss how BH spins can be 
constrained from the distribution of broad H$\beta$ line widths and optical 
luminosity (Fig. 5).  The sample of $z\lesssim 0.8$ quasars from Hu et al. 
(2008) appears to have moderate spins ($a \approx 0.5$), qualitatively consistent 
with the results from Wang et al. (2009), Li et al. (2012) and Volonteri et al. 
(2013) based on radiative efficiency considerations. Spin paramaters in the 
range $a=(0.5,0.7)$ correspond to $\eta=(0.08,0.1)$. 

High-luminosity quasars are normally difficult to study through RM because 
of their low level of variability, predicted long variability timescales, as 
well as $(1+z)$ time dilution (e.g., Kaspi et al. 2007).  If high-$z$ quasars 
follow the local size-luminosity relation, the H$\beta$ lags are predicted to 
be $380(1+z)L_{46}^{\gamma}$days, for $L_{46}=L_{5100}/10^{46}\,\ergs$.
From the results of this study, we note that the observed $L_{5100}$
is linearly proportional to $L_{\rm ion}$ only for $\bhm \lesssim10^8\,\sunm$. 
$L_{\rm ion}$ drops dramatically for BHs with very high mass and luminosity, 
which in turn leads to much shorter lags for the emission-line response.
Estimating from Figure 4, a $10^9\,\sunm$ quasar with $0.2L_{\rm Edd}$ will 
have its lag reduced to $\sim 20$, 50, and  200 days if $a=-1$, 0, and 0.998, 
respectively.  If BHs grow through random accretion (Wang et al. 2009; Li 
et al. 2012; Volonteri et al. 2013), some of them must undergo retrograde 
accretion and thus should have very short H$\beta$ lags.  This expectation can
be tested with future large-scale RM surveys.  Additional candidates of BH 
with retrograde spins can be selected from luminous quasars with very broad 
H$\beta$ profile in the FWHM$-L_{5100}$ plane (Fig. 5).

\acknowledgements The authors thank the anonymous referee for a helpful report
clarifying several points. We are grateful to the members of IHEP AGN group 
for stimulating discussions. This research is supported by the Strategic 
Priority Research Program $-$ The Emergence of Cosmological Structures of the 
Chinese Academy of Sciences, Grant No. XDB09000000, by NSFC grants 
NSFC-11173023, NSFC-11133006, and NSFC-11233003, and by Israel-China ISF-NSFC 
grant 83/13.  LCH receives support from the Kavli Foundation and Peking 
University.

\end{document}